\documentclass[a4paper,fleqn]{cas-dc}

\usepackage{amsmath,amssymb,amsfonts}
\usepackage{algorithmic}
\usepackage{graphicx}
\usepackage{textcomp}
\usepackage{tabu}
\usepackage{multirow}
\usepackage{booktabs}
\usepackage{makecell}
\usepackage{colortbl}
\usepackage{hhline}
\usepackage{hyperref}
\usepackage{rotating}
\usepackage{longtable}
\usepackage{comment}
\usepackage{enumerate}
\usepackage{enumitem}
\usepackage{multicol}
\usepackage{balance}
\usepackage{pdflscape}
\usepackage{afterpage}
\usepackage{rotating}
\usepackage[toc,page]{appendix}
\usepackage{float}
\usepackage[linewidth=1pt]{mdframed}
\usepackage{todonotes}
\usepackage{flafter}

\newcommand{\interviewquote}[2]{\begin{quote}
\small{\emph{``#1'' }} ---~\small{#2}
\end{quote} }

\usepackage{tikz}
\def\checkmark{\tikz\fill[scale=0.4](0,.35) -- (.25,0) -- (1,.7) -- (.25,.15) -- cycle;} 

\definecolor{Gray}{gray}{0.9}
\definecolor{LightCyan}{rgb}{0.88,1,1}
\definecolor{green}{rgb}{0.53, 0.66, 0.42}
\definecolor{itemgray}{rgb}{0.62,0.62,0.62}

\newcommand{\changed}[1]{\textcolor{black}{#1}}

\def\checkmark{\tikz\fill[scale=0.4](0,.35) -- (.25,0) -- (1,.7) -- (.25,.15) -- cycle;}

\def\BibTeX{{\rm B\kern-.05em{\sc i\kern-.025em b}\kern-.08em
    T\kern-.1667em\lower.7ex\hbox{E}\kern-.125emX}}

\usepackage[numbers]{natbib}
\newrobustcmd*{\citefirstlastauthor}{\AtNextCite{\DeclareNameAlias{labelname}{given-family}}\citeauthor}

\def\tsc#1{\csdef{#1}{\textsc{\lowercase{#1}}\xspace}}
\tsc{WGM}
\tsc{QE}

\ExplSyntaxOn
\keys_set:nn { stm / mktitle } { nologo }
\ExplSyntaxOff

\begin{document}

\shorttitle{Managing Security Evidence}

\shortauthors{Mazen Mohamad et~al.}

\title [mode = title]{Managing Security Evidence in Safety-Critical Organizations}

\author[1,2]{Mazen Mohamad}[orcid=0000-0002-3446-1265]
\ead{mazen.mohamad@gu.se}
\affiliation[1]{%
  organization={Chalmers University of Technology},
  country={Sweden}
}
\affiliation[2]{%
  organization={RISE Research Institutes of Sweden},
  country={Sweden}
}

\author[3]{Jan-Philipp Steghöfer}[orcid=0000-0003-1694-0972]
\ead{jan-philipp.steghoefer@xitaso.com}
\affiliation[3]{%
  organization={XITASO GmbH IT \& Software Solutions},
  city={Augsburg},
  country={Germany}
}

\author[1,5]{Eric Knauss}[orcid=0000-0002-6631-872X]
\ead{eric.knauss@gu.se}

\author[4]{Riccardo Scandariato}[orcid=0000-0003-3591-7671]
\ead{riccardo.scandariato@tuhh.de}
\affiliation[4]{%
  organization={Hamburg University of Technology},
  country={Germany}
}
\affiliation[5]{%
  organization={University of Gothenburg},
  country={Sweden}
}

\renewcommand{\shortauthors}{Mohamad et al.}

\renewcommand{\shorttitle}{Managing Security Evidence in Safety-Critical Organizations}

\begin{abstract}
With the increasing prevalence of open and connected products, cybersecurity has become a serious issue in safety-critical domains such as the automotive industry. 
As a result, regulatory bodies have become more stringent in their requirements for cybersecurity, necessitating security assurance for products developed in these domains. 
In response, companies have implemented new or modified processes to incorporate security into their product development lifecycle, resulting in a large amount of evidence being created to support claims about the achievement of a certain level of security. However, managing evidence is not a trivial task, particularly for complex products and systems.
This paper presents a qualitative interview study conducted in six companies on the maturity of managing security evidence in safety-critical organizations. 
We find that the current maturity of managing security evidence is insufficient for the increasing requirements set by certification authorities and standardization bodies. Organisations currently fail to identify relevant artifacts as security evidence and manage this evidence on an organizational level. One part of the reason are educational gaps, the other a lack of processes. The impact of AI on the management of security evidence is still an open question.

\end{abstract}



\maketitle

\section{Introduction}
Security is becoming increasingly critical in multiple domains, particularly for safety-critical systems \cite{securityImportant1}.
This is largely driven by the fact that products in these domains, such as automotive or medical devices, are becoming more open and connected, making them more vulnerable to cyber-attacks \cite{connectedVul1,connectedVul2,connectedVul3}. 
As a result, regulatory bodies have become more stringent in their requirements for cybersecurity, necessitating proof of cybersecurity for products developed in these domains, e.g., UNECE R156 regulation~\cite{unr156} in the automotive domain.
Standardization bodies have also begun issuing standards to ensure that companies in these domains employ security measures that meet the required levels to mitigate the high risks of cyber-attacks, e.g., ISO/SAE 21434 cybersecurity standard for road vehicles \cite{iso21434}.
Companies have responded to these regulations and standards by implementing new or modified processes to incorporate security into their product development life-cycle.
The goal of these efforts is to have security assurance which is defined as the ``Measure of confidence that the security features, practices, procedures, and architecture of an information system accurately mediates and enforces the security policy'' by the \citet{NIST}.

In security assurance, it is common to provide claims about the achievement of a level of security for a product or system \cite{haley2005arguing}. To support these claims, evidence has to be provided. Evidence in this context is an artifact that can be used to support or justify a claim about the security of the system. For example, test results of a security-relevant functionality, e.g., data encryption, a peer review on a design document, or a certificate of conformance with a standard. 
This evidence is also important in the process of cybersecurity certification, as it is used by assessors when evaluating certification requirements.

Managing evidence is not a trivial task especially when the developed systems are complex \cite{ruiz2011challenges}. This is due to factors such as a large number of evidence items, the complex organizational structures (including relations with suppliers), or the multitude and diversity of involved stakeholders.
Unlike in the field of safety, where safety assurance and evidence management have been studied extensively for a long time~\cite{nair2015}, evidence in security assurance has not been sufficiently covered in the literature \cite{SAC_SLR} despite its importance, especially compliance. A knowledge transfer from safety to security is possible. However, the differences between the two areas are important. For instance, security addresses malicious actors in the environment who impacting the system, whereas safety addresses accidental risks from the system that impact the environment~\cite{pietre2010sema}.
Additionally, there is a gap in the maturity of the two domains, especially with regards to standardization. Safety is considerably more mature and relevant safety standards have been introduced years ago~\cite{alexander2011}.

In this paper, we report on an empirical study based on interviews conducted at six different companies.
In the study, we aim at understanding the state of the practice with respect to working with evidence in organizations aiming at fulfilling the requirements of regulations and standards for security. 
We also investigate how the evidence is managed, what the main challenges in managing security evidence are, and to what extent automation can help.
Hence, we formulate our research questions as follows:

\leavevmode\newline
\textbf{RQ1. What is the context in safety-critical organizations with respect to managing security evidence?}

We understand \emph{context} as the factors that contribute to how organizations manage security evidence and the main drivers behind this. By understanding this context, we can investigate gaps in current practices. From this, we deduce the following sub-questions:
\begin{itemize}
    \item \textbf{RQ1.1} What is the level of maturity\footnote{The level of maturity refers to the extent to which processes, procedures, and technologies are established, standardized, and effectively implemented for evidence management.} with respect to managing security evidence in industry?
    \item \textbf{RQ1.2} What are the main drivers for creating and managing security evidence?
\end{itemize}
\leavevmode\newline
\textbf{RQ2. How is the management of security evidence embedded in an organization's development process?}

Similar to how safety evidence is handled, the management of security evidence needs to be part of the development process. To better understand how this happens in practice, we define the following sub-questions:
\begin{itemize}
    \item \textbf{RQ2.1} Which development artifacts serve as security evidence?
    \item \textbf{RQ2.2} Which activities\footnote{Activities refers to actions involved in collecting, storing, analyzing, or presenting security evidence.} during the development process create security evidence?
\end{itemize}

\leavevmode\newline
\textbf{RQ3. What are the detailed procedures aimed at managing security evidence in safety-critical organizations?}

While the results of RQ2 give us an overview of artifacts and activities, we explore the details of how security evidence is managed in this research question and its sub-questions:
\begin{itemize}
    \item \textbf{RQ3.1} Where does the responsibility of managing evidence lie?
    \item \textbf{RQ3.2} What processes are in place to manage evidence?
    \item \textbf{RQ3.3} How is access to the evidence handled?
    \item \textbf{RQ3.4} How is the evidence structured\footnote{Structure refers to how the evidence is organised and categorized.}?
    \item \textbf{RQ3.5} What practices exist to assure that the quality\footnote{Quality refers to the reliability, relevance, and credibility of the evidence to support a particular claim.} of the evidence is sufficient?
\end{itemize}

\leavevmode\newline
\textbf{RQ4. What are the challenges in managing security evidence for safety-critical organizations?}
We are aware that managing security evidence is a complex and challenging process. Hence, we aim to better understand these challenges in order to help organizations identify areas where improvements can be made.

\leavevmode\newline
\textbf{RQ5. To what extent can evidence management be supported through automation?}

Supporting security evidence management at scale requires automation of tasks during the development process. We explore this aspect with two sub-questions:
\begin{itemize}
    \item \textbf{RQ5.1} What is the state of practice on automation of tasks related to security evidence?
    \item \textbf{RQ5.2} What are the needs of the industry for the automation of tasks related to security evidence?
\end{itemize}

The contribution of this paper is an overview of the current state of managing security evidence. The study offers new insights into the ways in which organizations manage security evidence, including the integration of such evidence into the development process, detailed procedures for managing evidence in practice and the challenges associated with that, and the potential for automation to support evidence management. 
\section{Background and Related work}\label{sec:rw}

To the best of our knowledge, studies focusing mainly on the management of security evidence are scarcely reported in the literature. However, there have been related studies in connection with regulatory compliance including security. Additionally, there is related work that focuses on safety evidence management. Some of these studies target assurance evidence in general, and cover even security evidence. In this section, we give an overview of this related work.

\citet{Jaskolka2020} studies the challenges of providing security assurance for software-dependent systems by conducting a review of experience reports. The study lists multiple challenges related to working with evidence, e.g., coping with size and complexity of system which leads to a large amount of evidence required in the creation of security assurance cases. Another challenge mentioned in the study is dealing with external suppliers and the need to incorporate evidence from these suppliers into the assurance cases created internally. 
\citeauthor{Jaskolka2020} suggests recommendations for more effective security assurance solutions. These recommendations include, but are not limited to, developing tool support for integrating security activities into the software development life-cycle in order to provide sufficient evidence for security assurance and improving collaboration among stakeholders in the software development life-cycle to achieve better quality in terms of completeness of argumentation.

Another study by \citet{usman_rw} reports on an industrial case study that focuses on common practices and challenges with checking and analyzing regulatory compliance. The study is conducted at a large telecommunications company and contributes lists of challenges on regulatory compliance experienced by the company. Some of the challenges the authors identified, e.g., process-related ones, can be relevant in the context of security evidence. However, the paper makes no explicit reference to security evidence.

The study by \citet{beckers_rw} presents a method for establishing an information security management system (ISMS) that is compliant with the ISO~27001 standard. In the study, the researchers identify the different security artifacts (which we call security evidence in this study) required for conformance to the standard. However, the study does not propose management activities for these artifacts and rather focuses on the establishment and documentation of the ISMS. 

Automation of security evidence management has been studied by researchers. \citet{wali2013} studied building an automated security compliance tool that can be used by cloud providers to ensure a certain level of security for the enterprises using their services.
The tool includes an evidence engine that is responsible for creating storing and giving access to evidence when requested by users. 
A prototype of the tool was developed and integrated with a cloud platform to allow the users to automatically verify the status of compliance with the implemented security controls.

\citet{nair2015} study evidence management for compliance with safety standards.
The authors investigated the most frequently used safety evidence types in organizations and concluded that verification and validation (V\&V) artifacts, requirements specifications, and design specifications were the artifacts that were mostly used.
The study also highlights activities related to working with safety evidence which is usually manually performed in practice and would need automation and tool support, e.g., completeness checking and impact analysis. The study also shows that text-based techniques are used more frequently than graphical notations for evidence structuring and concludes that there is a gap between current research and industrial applicability.

Nair et al. \cite{nair2014extended} conducted a systematic literature review on provision of evidence for safety certification. As a result, the authors developed a classification of different artifacts that can be considered as safety evidence and review the techniques reported in the literature used for structuring the evidence. They found the most common structure is argumentation-induced and few papers use model-based specifications and textual templates. For the argumentation-induced structure, the authors list different notations  for structuring the evidence.
While \citet{nair2014extended} focuses on reported literature for safety evidence, our study is more practice-oriented and focuses on security evidence.

\citet{ruiz2011challenges} discuss the challenges of safety assurance and certification of safety-critical systems. The authors emphasize the high cost of safety compliance and that reusing safety arguments and evidence is a necessity to improve the certification process by making it more cost-effective and enable scalability. The study describes the need for an approach to specify, collect and manage safety evidence.

\citet{de2022model} suggest a model-based approach for managing assurance evidence for safety-critical systems. The study describes an assurance management process and provides a meta-model for assurance evidence. 
In this paper, we focus on security evidence management and we collect concrete data for aspects such as involved roles, data storage, and access control that can support organizations should they decide to adapt an approach similar to the one suggested by \citet{de2022model}.

Safety and security are related in multiple aspects. This has lead researchers to work towards approaches and methodologies for combined safety and security engineering.
In general, work related to safety and security co-engineering focuses primarily on safety and on security-informed safety analysis as emphasized by \citet{8556001} in their systematic literature reviews. 
However, for compliance reasons, there is a need to provide evidence for both aspects, which requires more studies on the security end, which we focus on in this study.

\citet{bramberger2020co} study the co-engineering of safety and security in the automotive domain. The authors highlight the lack of standards that provide structured co-engineering processes and that process developers need to consider and coordinate multiple standards to provide evidence of compliance to both areas. \citet{martin2017safety} also study co-engineering of safety and security and suggests an argumentation framework that enables product-specific safety and security combined analysis taking into consideration standard compliance processes.

\section{Research Methodology}\label{sec:method}
To answer our research quesions, we conducted interviews in six different companies. Some of the interviews were held with more than one interviewee. In this section, we describe our research method by providing an overview of the cases, our approach for collecting the data, and the methods we used to analyze it.
The authors followed the guidelines for qualitative surveys from the ACM Sigsoft Empirical Standards~\cite{ralph2021empirical}.

\subsection{Case companies}

\begin{table*}
\caption{List of the companies involved in the study.}
\label{tbl:companies}

\begin{center}
\begin{tabular}{@{}lllll@{}}
\toprule
             
            Company       & Location  & Size  & Domain & Role in value chain  \\ 

\midrule
             Case A     & Sweden  &     Large Enterprise           &   Automotive   & \changed{OEM}      \\            
             Case B     & Sweden  &     Small Enterprise           &   \changed{Automotive}  & \changed{Tool vendor}         \\
             Case C     & Sweden  &     Small Enterprise           &   Medical     & \changed{Service provider}      \\
             Case D     & Sweden  &     Large Enterprise           &   Automotive      & \changed{OEM}     \\
             Case E     & Austria  &    Large Enterprise           &   Automotive      & \changed{Component supplier}     \\
             Case F     & Germany  &    Medium Enterprise          &   \changed{Medical}    & \changed{Component supplier}       \\

\bottomrule
\end{tabular}
\end{center}
\footnotesize

\end{table*}

We conducted a study based on \changed{interviews} in six different case companies.
The companies were selected using convenience sampling out of our network of contacts with companies that build safety-critical systems that  also have a focus on security assurance.

In terms of size (number of employees), we included both large, medium, and \changed{small companies according to the definitions by the European Commission\footnote{\url{https://single-market-economy.ec.europa.eu/smes/sme-definition_en}}}.
All participating case companies are located in Europe. However, they all have global businesses stretching from Europe to America and Asia.
The participating case companies work in different domains, namely, automotive, medical, and software development. 
Table~\ref{tbl:companies} shows information about the case companies. The first column shows the identifiers of the companies, while the second shows the location of the company's headquarters. The third column indicates the sizes of the companies. The fourth column shows the domain in which the companies are active, \changed{and the last column shows the role of the company in the value chain}.

\emph{Case company A} is an Original Equipment Manufacturer (OEM) that produces passenger cars, while \emph{Case company B} is a tech company \changed{ and a tool vendor that produces solutions and system engineering tools in the automotive domain.} \emph{Case company C} is active in the medical domain and provides cloud services for patient data collection and monitoring. \emph{Case company D} is an OEM that produces heavy vehicles and trucks.
\emph{Case company E} is \changed{a very large} mobility technology companies for development, simulation, and testing in the automotive industry.
\emph{Case company F} is a software engineering company that builds high-end software systems for their customers, including software for medical devices.
All case companies provide products, services, and solutions in safety-critical domains. 

\subsection{Data collection}

\begin{table*}
\caption{List of interviewees. \changed{Participants from Case A and Case F were interviewed individually. In the rest of the cases, participants were interviewed as a group.}}
\label{tbl:interviewees}

\begin{center}
\begin{tabular}{@{}lllll@{}}
\toprule
             
            Participant   & Case company & Role & Years of Experience &  Security experience   \\ 

\midrule
             1     & Case A  &    Quality assurance expert    &   >20 & 2-5         \\
             2     & Case B  &    Solution manager    &   10-15   &  2-5         \\
             3     & Case B  &    Cyber security solution product owner    &  5-10 & 2-5         \\
             4     & Case C  &    Software development team leader    &   2-5 & 2-5         \\
             5     & Case C  &    Quality asset manager    &   5-10 & 0-2         \\
             6     & Case C  &    Quality assurance and regulatory affairs director    &   10-15 & 2-5         \\
             7     & Case D  &    Automotive connected solutions expert    &   >20 & 10-15         \\
             8     & Case D  &    Automotive connected solutions expert    &   10-15 & 5-10         \\
             9     & Case E  &    Technology scout    &   >20 & >20         \\
             10     & Case E  &    Lead engineer for cybersecurity    &   >20 & 10-15         \\
             11     & Case F  &    Head of IT    &   10-15 & 5-10         \\
             12     & Case F  &    Regulatory affairs expert    &   15-20 & 5-10         \\

\bottomrule
\end{tabular}
\end{center}
\footnotesize

\end{table*}

\changed{The data used in this study was collected by conducting interviews over a series of two-hour sessions at each of the participating case companies. In total, we held seven interview sessions (we did two sessions at one of the companies). We aimed at having multiple participants in each interview, but in some cases, we had only access to one practitioner. Hence, three of the interviews included two  practitioners, three had one (two of which were at the same case company), and one had three practitioners. The interviews were moderated by a single researcher, who was responsible for asking questions and leading the discussion, particularly in interviews with more than one interviewee which took on some characteristics of focus groups via discussion between the participants. The interviewer did take care to ensure all participants in group interviews were able to share their point of view, e.g., by specifically engaging interviewees who had not responded yet~\cite{smithson2000using}.}

Table~\ref{tbl:interviewees} shows the list of practitioners who participated in the study. The table shows the case companies to which the participants belong, their roles, their total years of experience, and their years of security-related experience.
As can be seen from the table, the majority of participants had over 10 years of experience in their domains. Also, the majority had multiple years of experience working with security. 

\changed{Identifying case companies and suitable interview partners proved to be a difficult endeavour. Security assurance is an emerging topic and not many organisations have the maturity to actively engage in topics of evidence management for security. We reached out to companies in Europe and the US and often received replies that indicated that the topic was on the agenda but that nobody worked on these topics at that point. This also makes it difficult to make an argument for saturation (see also Section~\ref{sec:threats}). However, since the aim of our research was to establish the current state of practice, we used the six companies that responded positively to our inquiries as a \emph{purposive sample}.}

\changed{All participants were provided with a consent form before the interview started. As is best practice, that form detailed the purpose of the study, anonymisation and data storage procedures, and the participants' ability to withdraw from the study at any time without negative consequences. All interviewees agreed to the conditions and returned the signed informed consent forms. The procedures outlined in the form were continuously monitored by the senior co-authors of this paper during the study.}

Before beginning the interviews, we provided a brief introduction to the study and conducted a round-the-table introduction of the participants. During this time, each participant shared their role, expertise, and years of experience, both in general and in their current position. This helped to establish a shared understanding of the group and set the stage for the discussion to come.

Once the introductions were completed, we began asking our pre-defined list of questions and initiated the discussions based on the focal points that we had identified in advance (see next sub-section). To capture the discussion in its entirety, all sessions were recorded with the participants' consent, which allowed the moderator to fully engage with the participants and ask follow-up questions without worrying about taking detailed notes. This method helped ensure that the discussion remained focused and that all relevant information was captured.

Once the sessions were over, we transcribed them and sent the transcription back to the participants in order to give them the chance to clarify any misunderstanding or correct any potential mistakes \changed{in a round of member checking~\cite{candela2019exploring}}. The participants did not request any changes.

\subsection{Focal points of the interviews}
\changed{The questions for the interviews along with the corresponding research questions are available in Appendix~\ref{appx:a}}. The questions are categorized into five different focal points.

\paragraph{State of practice}
In this point, we \changed{focused} on understanding the state of practice in security-related work at the company in question. We also \changed{asked} about security assurance and how it is addressed in the company and what the main drivers are for security assurance work.
Moreover, we \changed{asked} specifically about the security evidence to understand what types of evidence exist at the company, how they are produced, and when in the development life-cycle.
\changed{To support the interviewees in this point, we used the software security framework from the \emph{Building Security in Maturity Model} (BSIMM)  \cite{BSIMM} model that provides 12 different security practices divided into 4 practice domains.}

\paragraph{Responsibility}
In this focal point, we \changed{asked} about managing the evidence and particularly about the responsibilities of producing, maintaining, and collecting them.
We further \changed{asked} whether roles exist to manage and work with evidence and also about the main stakeholders and users of the evidence.
Additionally, we \changed{inquired} about potential stakeholders of the evidence and whether these stakeholders require special ways of managing the evidence.

\paragraph{Logistics}
For the logistics focal point, we \changed{asked} a number of questions to understand how the evidence is typically stored. We also \changed{inquired} about how access to the evidence is granted, such as through a secure login or other forms of authentication.  Additionally, we \changed{asked} about any processes that are in place for managing the evidence, as well as any recommendations for additional measures that should be implemented to better manage the evidence. 

\paragraph{Technical aspects}
In this focal point, we asked questions to gain a better understanding of why the evidence is collected, and how it is used. One of the questions \changed{was} about the motivation behind collecting the evidence, and we also ask for specific examples of how the evidence is used in different scenarios. 
Moreover, we \changed{asked} if there are types of evidence that are currently not being collected but would be necessary for current or future use cases.
We also \changed{focused} on the timing of when the evidence is created, such as during which activity or process step it's gathered. We also \changed{inquired} about the methods used to trace the evidence back to security issues or claims.
Additionally, we \changed{asked} about how the evidence is structured, e.g., on a product level, end-user function, etc. 
Lastly, we \changed{asked} how the evidence is maintained, the kind of properties it has, such as confidence or sensitivity, and how these properties affect or should affect the management of the evidence.

\paragraph{Automation} In this final point, we \changed{focused} on the automation of evidence management. In particular, we \changed{asked} whether there is any automation in place and if there are suggestions for the automation of tasks related to evidence management.

The \changed{interview guide is available in Appendix \ref{appx:a}} 

\subsection{Data Analysis}

\begin{figure}
    \centering
    \includegraphics[width=\columnwidth]{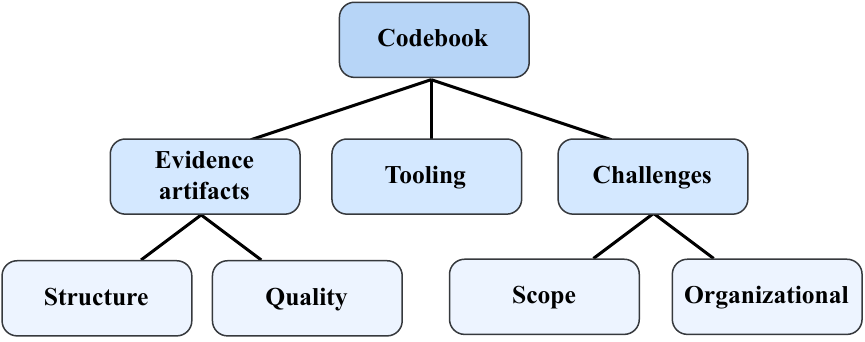}
    \caption{Excerpt from the preliminary codebook}
    \label{fig:codebook}
\end{figure}

To analyze and organize our data, we used thematic coding, which is a form of qualitative analysis that involves identifying texts that are linked by a common theme or idea and categorizing them \cite{gibbs2007}.
We applied the process described by \citet{clarke2015} and tailored it to our needs.
\changed{We also followed the advise by \citet{smithson2000using} to identify dominant voices or normative discourses in our data, to mitigate the limitations of the group interviews, but did not identify any such anti-patterns.}

\paragraph{Familiarization:} We \changed{familiarized} ourselves with the data by reading through it multiple times to gain a general understanding of the content.

\paragraph{Creating a codebook:} We created a preliminary codebook based on our understanding \changed{of} the context, the research questions, and the focal points we used in our focus group sessions. 
Figure \ref{fig:codebook} shows an excerpt of the preliminary codebook. As shown in the figure, we identified super-codes and then proceeded to create sub-codes as needed.

As an evaluation step, two researchers used the preliminary codebook to code one interview. Then a session was held to compare the codings and discuss the instances where they did not match. The codebook was refined based on the outcome of this activity. 

\paragraph{Conducting the coding:} We converted the transcripts into spreadsheets with individual rows containing the different statements by the interviewees and the interviewer. The spreadsheets also contained columns with the time stamps, for the codes, and for notes. 

We started the coding process by reading the transcripts line by line categorizing the text chunks from the focus groups and labeling it with our pre-defined codes.
During this process, new codes emerged. We merged those to our codebook after revising them. Additionally, some pre-defined codes were split or merged as we learned more about the data.
The final codebook consists of 63 codes grouped into 9 super-codes. It is available online \cite{supplemental-material}.

To illustrate our coding process, we give the following example:

\interviewquote{That end-user-function can cross so many different domains, it can be people, it can be processes, technology, onboard, offboard, communication, electrical components}{Case D}

The statement talks about a challenge associated with the complexity of end-user functions in a vehicle. Hence, we used our super-code ``challenge''. However, the code is too abstract, so we used our sub-code ``scope'' for that statement.

\paragraph{Searching for patterns:} When the coding was done, we conducted two workshops to analyze the coded statements and look for patterns. 
The workshops were conducted by three researchers each, and all authors of this study participated in at least one workshop. 
We clustered similar statements into themes based on the main ideas they include.
We used an online tool \cite{miro} to illustrate the statements and codes, and reshuffle and group them into their respective themes.
We also linked these clusters to our research questions. As a result, we identified 56 clusters in total.

\paragraph{Identifying findings:} Finally, we used the themes and codes to draw insights and conclusions from the data.

\section{Results}\label{sec:res}
In this section, we present the results of the study structured by research question. 
\changed{We trace the results to the cases by specifying the corresponding cases within brackets using the case identifier used in Table \ref{tbl:companies}}
\subsection{RQ1: What is the context in safety-critical organizations \changed{with respect to} managing security evidence?}

\subsubsection{What is the level of maturity \changed{with respect to} managing security evidence in the industry?}
\vspace{5pt}
\changed{Our results indicate that there is awareness of the importance of security evidence management, but despite that, there is immaturity when it comes to the management of security in general and security evidence in particular [A,D,F].}

\changed{\interviewquote{The automotive industry as such is a bit immature when it comes to security} {Case~D}}

Additionally, our results show that there are varying degrees of maturity within the case companies, particularly in larger organizations \changed{ [A,D,E,F]}. This is due to the fact that different teams within these organizations may use different technologies with varying levels of security, \changed{and also have different security competencies among the team members and different ways to manage security evidence.}
\interviewquote{It’s different maturity levels in different teams at [the company] as we work in different environments. Some work in cloud environments, others work in embedded hardware, and we have different ways of working with this.}{Case~D}

\changed{Four of the case companies [A,D,E,F]} are in the stage of establishing processes and templates for working with security in order to ensure consistency among the teams and avoid potential security gaps.
\interviewquote{We are still in the discovery phase, but I think the best thing is to invest time so that the team has a unified opinion because there are often problems with one term and two people understanding it completely differently.}{Case~F}

Knowledge transfer from the safety domain is a possibility according to the participants of two \changed{interviews }\changed{[A,D]}. However, it is essential to note that security and safety are distinct domains with different needs and priorities and this must be considered when transferring knowledge. For example, security, in contrast to safety, has to deal with intelligent agents that have the intention to cause harm to a system \changed{\cite{pietre2010sema}}. This causes a high level of uncertainty about attackers’ behavior. Hence taking measures that are not responses to specific threats is a common practice \changed{\cite{alexander2011}}. \changed{Additionally, recent security standards require continuous risk assessment, which might lead to new evidence created during a product's life-cycle \cite{iso21434}. This implies that a security body of evidence is more dynamic compared to a safety one and needs to be managed accordingly.}

\subsubsection{What are the main drivers for creating and managing security evidence?}
        
            All participants emphasized the growing importance of security in their respective fields.  One important aspect is \changed{providing security evidence of products and services supplied by the companies [B,C,D]}. Another aspect that was mentioned in two of the \changed{interviews [B,D]} is the market for security services \changed{including evidence management}, i.e., providing security solutions and consultation services to customers. 
            
            The interviews revealed several drivers for working with security evidence. Compliance with standards and regulatory requirements, such as UN-r155 \cite{unr155} \changed{[D,E]}, ISO 27000 \cite{iso27000} \changed{ [A,C]}, ISO 27001 \cite{iso27001} \changed{ [A,C,D,F]}, HIPAA \cite{hipaa} \changed{ [C,F]}, GDPR \cite{gdpr} \changed{[C]}, and ISO/SAE 21434 \cite{iso21434} \changed{[A,B,D,E]}, was identified as a significant driver, as the evidence would be used to prove compliance. 
             
            Customer demands and requirements also played a major role, as failing to meet these could result in a loss of customers and economic damage to companies: 
            \interviewquote{Actually some customers started to put demands on us to upgrade the security of our system to match their internal policies, and failing to do so from our side would make us lose that customer}{Case B} 
            Hence, evidence can play a major role to give customers a level of confidence in the security of the provided products, services, and systems.

            Internally, the participants emphasized their recognition of the need for confidence in the level of security and the potential opportunity to enter the security market as a driver \changed{from a management point of view}. 
            \changed{Especially with the increasing technological complexity of today's products, such as vehicles and medical equipment, where the risk of cyber-attacks also rises.} 
            \changed{\interviewquote{Market part is a big opportunity from a management perspective, and resources are being employed to get into the market}{Case E}}
            \vspace{-15pt}
            \changed{\interviewquote{the majority of the cars nowadays are connected which makes them vulnerable to outside attacks, especially that most mechanical functionalities are now becoming software-driven}{Case B}}
             
            Additionally, the threat of litigation was considered a driver for security assurance and evidence collection:
            \interviewquote{If someone finds out a security flaw, there might be a follow-up asking: you told us this is secure, then the evidence might be employed to show what has been done.}{Case E} 
             
            Overall, the participants highlighted the need for their respective businesses to prioritize security in order to protect themselves and their customers in an increasingly risky environment and to comply with current and upcoming standards and regulations, and consider evidence to play a major role in that.

\subsection{RQ2: How is the management of security evidence embedded in the development process of an organization?}
        Various activities are carried out throughout the development life-cycle to create security artifacts.
        The participants mentioned various types of evidence created through these activities.

        In the early process stages, activities such as \emph{asset identification }\changed{ [C,D,E]}, \emph{risk analysis} \changed{ [A,C,D,E]}, and \emph{gap analysis} \changed{ [F]} are performed to identify potential risks, evaluate them, and propose mitigation strategies.
        \interviewquote{Risk analysis evaluates the risks and then proposes mitigations, and then we implement those and those can affect the product where they could affect that the infrastructure or some process.}{Case C}
        These activities result in artifacts used in the argumentation about the security of the system, as well as evidence created to prove that the chosen mechanisms work properly.
        \emph{Risk analysis}, considered one of the most important activities by the \changed{interviewees}, provides the basis for the argumentation part of an assurance case. It is a crucial activity that produces evidence. Additionally, it produces assumptions that can be used in the arguments made about the security of the systems \changed{[D]}.

        \emph{Security overview documents}, such as scorecards and compliance matrices, are presented internally to higher management and provide evidence of requirements fulfillment and degree of compliance \changed{[A]}. 
        
        Verification activities, such as penetration testing, code reviews, and dynamic analysis, produce \emph{test reports} that are also considered significant evidence \changed{[A-F]}.

        \emph{Policies} \changed{[C,E,F]} and \emph{process descriptions} \changed{[F]}, as well as \emph{security training and awareness programs} \changed{[B-D,F]}, are considered evidence by some companies. Two case companies have defined processes for \emph{training}, e.g., with respect to the frequency and level of training \changed{[C,F]}. Having such a process and being able to prove that the employees participate in it might be used to justify claims about security, as it enhances the secure development competencies of developers, creates a security culture, and raises awareness about security among other roles in the organization. This can be used as evidence to raise the confidence level in the security of the products these companies provide to their customers.

        \emph{Logs} related to incidents can be used as evidence for periodic checks, such as access rights \changed{[C,F]}. 
        The architectural and technical design of the system, including the \emph{security assessment} of cloud providers or secure technologies, are also sources that provide evidence \changed{[B,D]}.
        Additionally, \emph{Personnel background checks} are also conducted to mitigate the risks of internal breaches \changed{[D]}.

        \emph{Governance} activities contribute to the creation of security artifacts by providing \emph{guidelines and best practice documents}, developing \emph{cybersecurity plans}, and performing \emph{screening activities} for design, conceptualization, implementation, and validation and verification \changed{[D,E,F]}. 
        
        On the IT security side, activities such as \emph{attack simulation}, \emph{system monitoring}, \emph{virus scanning}, \emph{penetration testing}, creating \emph{sandboxes}, and \emph{security assurance cases} are performed to assess \changed{and provide evidence of} the level of security for the system in question \changed{[B-D,F]}.

        Finally, late-stage development activities, such as \emph{deployment processes}, \emph{automatic testing}, \emph{joint reviews}, and \emph{approval with suppliers}, also result in evidence \changed{[A,C,E]}.
        
        \changed{Overall, there are various types of activities and resulting evidence that can be used for ensuring the security of an organization's products or services. By using these types of evidence, companies can make more informed decisions and effectively demonstrate their security posture to stakeholders. However, the results we report here are not meant to be a complete list of evidence, but rather an overview of the different types of evidence created by the companies in the current state.}

\subsection{RQ3: What are the detailed procedures aimed at managing security evidence in safety-critical organizations?}

\subsubsection{Where does the responsibility of managing evidence lie?}
        \begin{table*}
\caption{Roles and responsibilities of managing security evidence}
\label{tbl:res:resp}
\begin{center}
\begin{tabular}{lccccc}
\toprule
             
            \textbf{Role} & \textbf{Creation}       & \textbf{Ownership}  &   \textbf{Collection}  &   \textbf{Maintenance}  & \textbf{Governance}  \\ 

\midrule
         Developer / DevOps   & \checkmark     &  &                &   \checkmark &        \\
         Product owner   &      &  \checkmark &    &   &    \checkmark     \\
         Risk owner   &     &   \checkmark &      & &         \\
         Auditor   &      &  &       &  & \checkmark          \\
         Management   &      &   &       & &  \checkmark       \\
         Security officer   & \checkmark   &   & \checkmark &   \checkmark   &\checkmark        \\
         Legal team & &   & \checkmark & & \\
             
\bottomrule
\end{tabular}
\end{center}
\footnotesize
\end{table*}

\emph{Developers} have been primarily responsible for creating and maintaining evidence \changed{[C-F]}. However, this is changing as the scope and complexity of evidence management increase \changed{[C]}. While developers will continue to be responsible for creating and maintaining evidence, other specialized roles are taking on additional responsibilities for managing evidence, including ownership and governance.
\interviewquote{Working with evidence has been up to the development team until recently, but there is work that is driving this to be formal. If it is related to product, it is the development teams. There are different roles in the teams and they produce different evidence related to their work.
There is also the role of quality assets manager related to the system management, and we have a role called privacy and security officer. It is a high level role and it’s related to regulations, e.g., GDPR and HIPAA when we need to have external communications to external parties.}{Case C}

\emph{Auditors} and \emph{compliance personnel} have a role in ensuring that evidence is being collected \changed{[D]}. They also have the responsibility to make sure that the evidence meets the required quality levels and is reliable. 

\emph{Product owners} act as owners of the produced artifacts, including evidence. They have the responsibility of ensuring that the evidence is being created and maintained. Additionally, they have the responsibility of providing access to the evidence to potential stakeholders \changed{[A]}.

In organizations where the role of \emph{risk owner} exists, that role acts as the owner of the evidence associated with the risks they own \changed{[D,F]}. There may be some overlap in ownership between the risk owner and the product owner, but both are responsible for governing the evidence by ensuring that it is created and maintained.

Management, which may include \emph{project managers}, \emph{line managers}, \emph{product managers}, and others, has the primary responsibility for governing the evidence work by ensuring that the security policies and processes are in place and functioning properly \changed{[B-F]}. Although not a direct user of the evidence, management is concerned about gaining an overview of the security status of their products and systems \changed{[B]}. 
\interviewquote{The upper management would be interested. However, the lower level technical details are up to the team, but the high level would be interesting for the top management to be able to gain their trust in the team.}{Case B}

\emph{Security officer} is another role mentioned in three \changed{interviews} \changed{[A,E,F]}. The responsibility of security officers is similar to management, but they have more involvement in the creation, collection, and maintenance of evidence. The security officers work more closely with the teams to request evidence. Additionally, the evidence which is of interest to the security officers is usually on a more detailed level than what management requires, as management is usually interested in a more abstract view of the security status \changed{[D]}.

The \emph{legal team} has been mentioned in one of the \changed{interviews} \changed{[D]} as being responsible for the collection of evidence. The driving factor is when claims and law suites are filed against the case company. In that case, the legal team would be collecting all evidence connected to the case, e.g., evidence about the security of a certain module in the product that caused the incident. This evidence would be presented to the court, and can also be used by the opposing party. Hence, the evidence needs to be of good quality. 

Finally, other roles are also mentioned, such as \emph{asset managers} \changed{[C]}, \emph{privacy and security officers} \changed{[C]}, \emph{chief information service officer} \changed{[A]}, and \emph{project security manager} \changed{[E]}. These roles have similar responsibilities to the security officer role.

\changed{Based on these results, we identify five main responsibilities for managing security evidence. These are: \emph{(i)} creation of evidence; \emph{(ii)} ownership of the evidence; \emph{(iii)} collecting the evidence; \emph{(iv)} maintaining the evidence; and \emph{(v)} governance of the evidence.
Table~\ref{tbl:res:resp} shows these responsibilities and the corresponding roles carry them out.}

\subsubsection{What are the processes in place to manage evidence?}

During the \changed{interviews}, several important aspects related to managing evidence were identified. Firstly, \emph{risk management} was found to be a common practice, and the most relevant step in this process was considered to be risk analysis, \changed{as it is an important step in deciding what evidence is required and how to manage it} \changed{[A,C,D]}.
However, conducting risk analysis on complex infrastructure and microservices can be challenging, and identifying patterns of risk analysis on a component level could be used to cope with this, as it enables reusing parts of the analysis for similar components \changed{[D]}. 
\interviewquote{We have our infrastructure, microservices, number of services, and architecture where everything is talking to everything. It’s extremely difficult to do a risk analysis on them. So what we have done is that we have defined how we build a component and all the surrounding infrastructure and dependencies and how we operate that component. And then we have performed kind of a pattern risk analysis on that component.}{Case~D}

\emph{Traceability} was also highlighted as an essential aspect of managing evidence \changed{[A-F]}. Although the state of practice varied among different case companies, traceability was considered important, especially because it was required by regulations \changed{[D]}. Traceability needs to connect the development artifacts on multiple levels, from the requirements to the code and testing artifacts \changed{[C]}. It provides a link between the security arguments derived from requirements and architecture and the evidence, enabling security assurance \changed{[B]}.

\emph{Change management and maintenance} were found to be processes related to risk analysis, as they handle how risks are updated and maintained in case of changes \changed{[C-E], which might affect security evidence related to those risks}.
\interviewquote{We have change management as a process, if it's related to risks, then the risk analysis should be updated. It might lead to new mitigation and then that would end up in a Sprint planning since we drive development with sprints, and there would be implemented.}{Case~ C} 
Incident management was another process that needed to be implemented similarly to handling any discovered bug \changed{that might affect certain evidence or even require new evidence for new risks} \changed{[A,E]}.

\emph{Policies and checklists} were commonly used to manage security \changed{in general, and security evidence in particular}, with different companies having different policies in place \changed{[C,F]}. Checklists are used to ensure the completeness of certain tasks, \changed{e.g., creation of evidence,} and could be included in more generic documents, such as the Definition of Done \changed{[F]}. 

Processes for \emph{exchanging information} are also important \changed{[D]}, and ownership of evidence and security artifacts needed to be determined when exchanging information between organizations, teams, or systems.

Lastly, \emph{requirements management} is mainly used when working with suppliers, allowing the companies to impose requirements on the suppliers to use certain technologies or functions \changed{and request evidence to verify the fulfillment of these requirements [D]}. This is done to ensure that the companies are accountable for their products, regardless of the suppliers they worked with: 
\interviewquote{[The company] is responsible towards their customers, and can never blame the supplier, so we need to work closely with the suppliers and put requirements to specific technologies or functions delivered by them.}{Case~D}

\subsubsection{How is the evidence stored and how is access to the evidence handled?}

\begin{table*}
\caption{Approaches of storage and access control of security evidence according to the \changed{interviewees}}
\label{tbl:res:storage}
\begin{center}
\begin{tabular}{lll}
\toprule
             
            \textbf{Approach}  & \textbf{Advantages}       & \textbf{Disadvantages} \\ 

\midrule
         Centralized &  Easy access control across systems   &  Hard to achieve    \\
                     &  Enables re-usability    & Hard to maintain     \\
                     &                         & Elevated privileges for administrators     \\
                     &                         & Imposes security risks     \\
\midrule
         Decentralized   &     Easy access control within a system &  Hard to collect evidence across systems      \\                         
                         &     No overhead of transferring evidence & Hard to conduct quality control      \\
                         &     Can be managed by different teams   &        \\             
\bottomrule
\end{tabular}
\end{center}
\footnotesize
\end{table*}

We discussed the methods used for storing and accessing evidence for their various projects and systems in the \changed{interviews}. 

Two main approaches emerged during the discussion. They are \emph{(i)} using centralized storage for the evidence and hence having a centralized access control mechanism; and \emph{(ii)} decentralized storage where the evidence is stored in repositories specifically designated for each project or system and accessed through access control mechanisms available for these systems. 
Table \ref{tbl:res:storage} shows a summary of the advantages and disadvantages of the two main approaches for storing and accessing evidence.

Two case companies used the centralized approach by utilizing specialized tools for storing and accessing evidence \changed{[B,E]}, with one tool originally designed for functional safety, but now also utilized for cybersecurity \changed{[E]}. Access to the evidence is granted through the tool itself, making it easier to manage and control access across multiple systems. Moreover, storing evidence belonging to the same architectural domains centrally is beneficial for reuse purposes and follow-up activities such as vulnerability management when required \changed{[E]}.  However, centralized storage is very difficult to achieve in big organizations with complex products \changed{[A,D]}. Moreover, centralized storage systems can give administrators complete control over the data, which can create concerns around privacy and data ownership \changed{[D]}. Additionally, storing all the evidence in one place is not considered very secure, as it could include sensitive data and would be a potential target for cyber-attacks.

\interviewquote{I would say from a least privileged access principle, it is not good to store all of this evidence in one place. Because this evidence,  even though it's proof that something is secure, is an attractive material for attackers.}{Case~D}

The decentralized approach which is used by the remaining case companies \changed{[A,C,D,F]} is adopted by storing evidence either according to the product or project structure \changed{[A]} or according to the different teams in the organization \changed{[D]}. This means that access to evidence is based on the responsibilities of each team member, as set up in the project structure. By following this method, it becomes easier to manage and control access to evidence \changed{[D]}.
Additionally, other evidence types can be stored in tools such as version control tools \changed{[C,F]}, wiki pages \changed{[F]}, and bug trackers \changed{[F]}. The users then use the functionalities of these tools to control access to the evidence.

However, a challenge that emerges is the need for stakeholders to access evidence of multiple systems, sub-systems, or modules. For example, security assurance for certain functionality in a vehicle requires collecting evidence from multiple modules, embedded systems, and back-end systems \cite{mohamad2020security}. In such cases, access needs to be granted to all relevant systems, which could prove cumbersome to manage. This issue can be mitigated by having a specialized tool that provides trace links to the actual evidence \changed{[A,C,D]}. Hence, the tool would not include the actual evidence but rather links to where this evidence is stored. Another disadvantage of this approach is that applying processes for, e.g., quality control of the evidence becomes hard when the evidence is scattered \changed{[A]}.

A way to mitigate the disadvantages of having either approach and utilizing the advantages is to apply a hybrid approach as discussed by the participants of one case company \changed{[A]}. In this approach, a subset of the evidence is stored centrally, but not all of it. This does not entirely eliminate the disadvantages of the two approaches but rather helps mitigate them. For example, a company can identify critical types of evidence that are frequently collected for different usages and decide to store them in shared storage, which makes accessing those easier. However, in a scenario where there is a need to exhaustively collect evidence for a certain product/system, then the disadvantage of scattered evidence locations persists.

\subsubsection{How is the evidence structured?}

The participants discussed various approaches to structuring security evidence. participants from three case companies \changed{[B,D,E]} discussed that depending on the project or product in question, evidence may be structured as feature-based or architecture-focused:

\interviewquote{In many cases, the project is feature-based, hence the evidence is also feature-based.}{Case~E}

The evidence needs to be linked to the product structure in a way that allows for easy implementation and tracking of tasks and issues \changed{[B]}. To achieve this, evidence needs to be organized in a flexible and responsive way that meets the demands of external entities such as OEMs or Tier-1 suppliers.

It is important that the evidence is not managed in silos but rather through the same working methods as the rest of the development process \changed{[E]}. This makes it easier for implementation artifacts to refer to the evidence and identify and address security-related issues.

\changed{Participants in three interviews} \changed{[B,D,E]} mentioned that no specific structure is in place to store security-related evidence in their organizations or sub-organizations. However, there are attempts to form a plan for structuring evidence. One proposed approach is to create a map or index of evidence to help specific roles locate evidence related to their function, providing a clear overview of where the evidence is located rather than sorting it in a particular way \changed{[D]}.

Another point mentioned in one \changed{interview} \changed{[A]} is that the evidence needs to follow the quality assurance setup that already exists in a company rather than creating a special one. This is mainly because security evidence should not be viewed as unique artifacts according to the participants.

\changed{According to two interviews [A,C], tools rather than evidence storage are the main drivers for how the evidence is structured}. \interviewquote{The used tools drive the structure of the evidence. It's not how we store the evidence that drives the structure but rather how the tools work and how we use them.}{Case~C}

Finally, one approach to structuring evidence is based on models required or recommended by standards \changed{[A]}. In this approach, evidence would be structured based on the requirements of the standards. For example, if a standard requires a threat analysis to be performed on a module level, then the evidence would be structured on the same level. 

Overall, the \changed{interviewees} emphasized the importance of flexibility and responsiveness in structuring security evidence, as well as the need for clear organization and accessibility.

\subsubsection{What practices and measures exist to ensure that the quality of the evidence is sufficient?}

The quality of evidence can be assured through the application of a set of activities and procedures. Different measures can be used to assess the quality level of security both on an organizational level and also on a product level. 

Organizational level activities pointed out by participants from three case companies \changed{[A,D,F]} include the use of an \emph{industry-standard guideline} for evaluating software development processes, e.g., the Software Process Improvement Capability dEtermination (ASPICE) in automotive \changed{[A]}, or through \emph{audits} performed on suppliers to assess their security level \changed{[D]}.
In some cases, \emph{joint risk analysis} can be performed with suppliers to increase confidence in the applied risk treatments. The metrics that result from these activities are usually also on an organizational level \changed{[D]}. \changed{These activities can be used to assess the quality of evidence. For instance, during the combined risk analysis, evidence can be presented and evaluated.}

\changed{\interviewquote{We audit the suppliers and collect evidence that way. For instance if a partner is developing a software for back-office systems, we do an assessment of that partner and we can ask for audit reports, like if they have ISO 27001 certification. We do other types of assessments, like performing risk analysis together with their development teams. We review their risks analysis and manage risks together with them. So in one way we could think of them as being a sub-part of our organisation. That type of evidence is really important.}{Case D}}

When it comes to metrics on a more granular level, \emph{test coverage} is a common one \changed{[B,E]}. It is used to show confidence in tests \changed{(which are usually used as evidence)} and is often a requirement in certain standards.

\changed{\interviewquote{On the testing side, test coverage can be a metric which is measurable and can show confidence. It is also required in the ISO [ISO/SAE 21434] standard}{Case E}}

Risk analysis is also quality assured using metrics such as \emph{completeness} and \emph{correctness} \changed{[D]}.
Another activity is to consider a \emph{list of threats} derived from regulations to ensure an acceptable level of \emph{threat coverage}. This can be used to calculate a metric for threat coverage \changed{and used as a piece of evidence.} \changed{[E]}.

One participant \changed{[A]} stressed that in project-based organizations, the process of quality assuring the evidence should follow the general quality assurance procedures of the project, and the responsibility for assessing security artifacts and the evidence lies with the project team. 

\emph{Information filtering} is another important step in \changed{evidence quality assurance}, where properties are assigned to the information \changed{[B,C]}. This is done by tagging certain artifacts such as test cases, requirements, and implementation tasks as security-related. A \emph{sensitivity level} can also be assigned to different \changed{evidence} to decide how they should be \changed{managed}. Participants from \changed {two interviews [B,C] identified these properties as important aspects that need to be implemented}. Additionally, they suggested assigning other potential properties to evidence such as \emph{confidence in evidence coverage}, \emph{confidence in the people producing evidence}, and the \emph{organization's cybersecurity awareness level} \changed{to help assessing its quality}.

It is noteworthy that our interviewees mostly mentioned techniques to \emph{assess} the quality of evidence, but did not provide information about more constructive methods to assure that evidence quality is high.

\subsection{RQ4: What are the challenges in managing security evidence for safety-critical organizations?}
Managing security evidence is a complex task that poses numerous challenges for organizations. These challenges were discussed in our \changed{interviews}, where the participants shared their experiences and insights on the matter.

One of the most significant challenges mentioned by several participants was establishing the required scope of working with security evidence \changed{[A,D-F]}. This challenge arises due to the uncertainty and constantly changing nature of security, where new threats and vulnerabilities are discovered regularly \changed{[A,D]}. It also becomes impossible to cover 100\% of cases, which makes the aim to create a \emph{secure enough product} rather than a secure product \changed{[D]}. Moreover, the complexity of products and systems with a high number of dependencies among them makes setting a scope for security a challenge. This has a direct effect on the evidence, as it becomes important to determine the needed evidence to make sure that the produced products are secure enough.

The diversity of products and organizations also makes it challenging to define methodologies and processes that fit all \changed{[A-F]}. Security processes need to handle the rapid growth in the market and the growing complexity of product development, which makes the scope of security assurance and evidence wider with time \changed{[F]}. Companies need to find a balance between their specific security posture and requirements and the standardized approaches required by regulations \changed{[E,F]}.

Another significant challenge is sharing sensitive data \changed{[B]}. Security evidence has the nature of being sensitive, and this creates challenges regarding sharing this data both internally within a company and externally with third parties, e.g., suppliers, customers, and auditors. 
As security awareness improves and policies are applied, the challenge of sharing sensitive data increases. \interviewquote{The old way of accepting that a lot of data is open to everyone and trust will change.}{Case B} This can lead to increased bureaucracy and delays in the exchange of data \changed{[B]}.

The security demands require \changed{providing and managing evidence for} the complete life cycle, which is a challenge in big organizations, but more so with supplier-customer relationships \changed{[A,D]}. \interviewquote{Right now the supply chain needs to be able to provide inside knowledge and support throughout the lifecycle which was not really [necessary] before.}{Case A} Security needs to be addressed throughout the product's life cycle, including its development, deployment, operation and decommissioning \changed{[D]}. The challenge of covering the full life cycle is that the decommissioning stage is usually forgotten. Companies need to consider the security of their products beyond their immediate use to prevent any potential vulnerabilities from being exploited after the product has been retired. \interviewquote{You know we create something secure in a vehicle and the vehicle has a life-cycle of 15-20 years, then how secure is it when this time has passed?}{Case~D} Hence, they need to create evidence at design and implementation time that covers the full-lifecycle, which is considered very challenging according to our results.

The cost and effort of working with security evidence can be considerable, and estimating the return on resources invested on security is a challenge \changed{[C,D]}. It is also challenging to determine the optimum level of security in relation to the cost and effort invested. \interviewquote{There is always a breaking point where you spend too much time, resources, and money to get those extra last miles of security, which doesn't really give anything. It wastes a lot of energy in the company, so you need to figure out where the optimum level is here.}{Case~D}

Working with security evidence requires collaboration among different teams, but there is often a lack of competencies when it comes to security, making it challenging to establish processes and policies \changed{for that purpose. [A,D-F]} \interviewquote{There is totally a war out there for security competencies.}{Case~D}

Despite having standards and regulations with requirements on security, there is a lack of details in these documents on how to manage and work with security evidence. It is up for interpretation in many cases, making it difficult to establish a standardized way of working with security evidence \changed{[F]}. \interviewquote{ISO 9001 has very vague requirements on information security. Nothing concrete.}{Case F}

Identifying security relevance \changed{where evidence is required} is also challenging, from interpreting requirements to developing new features and making changes to systems. Companies need to interpret these requirements and regulations to suit their specific needs \changed{[C]}. \interviewquote{Also it would be interesting to have a process to assess the importance of a change and how relevant it is for security.}{Case C''}

\subsection{RQ5: To what extent can evidence management be supported through automation?}

\subsubsection{What is the state of practice on automation of tasks related to security evidence?}
The participants from all case companies agree that automation of tasks related to evidence management can be a significant benefit for organizations. 
Some automation of tasks is created in the current state of practice. 
One critical component mentioned \changed{in three interviews} \changed{[D-F]} is \emph{report generation}, which can include creating reports based on evidence requirements from standards and regulations, logging reports, build reports including testing and scanning reports, and reports generated from wiki pages for specific items, e.g., a certain process.

Another critical component for automation is \emph{anomaly detection and monitoring} \changed{[C,D,F]}, which is considered essential for cybersecurity \changed{and an important source for evidence creation}. Automation can help organizations quickly identify and respond to threats, and monitoring and detection mechanisms can be used to understand the attractiveness of systems to attackers \changed{[D]}. \changed{Cutting-edge technology such as machine learning can also be used for anomaly detection and monitoring [D]}. Automated monitoring, alerts, and logging can help identify anomalous behavior, and traceability of changes in code can be used to quickly identify potential threats \changed{[C,F]}. While intrusion prevention is not always possible, intrusion detection in the network is still considered essential \changed{[F]}.

\emph{Testing and verification} \changed{ where test reports (a major type of evidence) are created,} are also important components of automation in evidence management. One participant \changed{[E]} mentioned using \emph{automated testing} and \emph{model-based testing} for their products, with automated test cases being created based on threat models. Tools are also used for code reviews that enable \emph{automatic monitoring and documentation of tasks and responsibilities} \changed{[C]}.

Finally, one participant \changed{[E]} emphasized the importance of automation in \emph{producing attack models and trees}, allowing engineers to focus on modeling and designing systems and defining the right attributes. Automation in threat modeling can save time and effort and reduce the risk of human error. \interviewquote{In the Threat Analysis and Risk Assessment (TARA), there are efforts to more and more introduce automation, so the engineers can focus on modeling and designing these systems and defining the right attributes, but the threat models and attack trees are automatically created.}{Case E}

\subsubsection{What are the needs of industry for automation of tasks related to security evidence?}

The \changed{interviewees} have identified several areas where automation can improve security practices \changed{in general and evidence management in particular}. One such area is \emph{security relevance prediction} \changed{[B,C,E]}. It is very important for practitioners to be able to identify the security relevance of various artifacts \changed{(including evidence)}, such as requirements, test cases, implementation tasks, etc. This allows them to apply the appropriate processes and management procedures to these artifacts. 
This task is easier for new development projects than for deployed systems. 
For example, when arguing about the level of security for a deployed system, the practitioners must be able to identify those requirements that are security-relevant to gain knowledge about what has been developed in that context. \changed{These arguments would then be justified using evidence.} The same thing applies to identifying assets. This task becomes easier if the requirements are labeled as security-relevant/non-security-relevant, but that labeling is often not available for deployed systems. 
Another example is the labeling of test cases that potentially could be used as evidence. To overcome this issue, a participant \changed{[B]} suggest using machine learning techniques to predict the relevance of these artifacts to security.

Another important area where automation can be useful is in \emph{test case generation and selection} \changed{as the results of test cases is commonly used as evidence} \changed{[C,E]}. Automatic generation of test cases for security-related development tasks can be used in combination with threat modeling tools that automatically yield threats \changed{[E]}. Test selection is also important, allowing the selection of which test cases to run based on the changes made to the code. Hence, if a change concerns security, then those test cases that cover the security of the system shall automatically run \changed{[C]}.

\emph{Traceability} is another area where automation can be highly beneficial. Interviewees emphasized the need for traceability in effective risk handling and evidence creation \changed{[A-C,E]}. 
One participant \changed{[A]} stressed the importance of combining traceability with the TARA process. This would enable developers to identify potential risks and vulnerabilities and take steps to address them before they become major issues. 
 Another participant \changed{[B]} emphasized the need for \emph{automation in the change impact analysis} to detect the impact on security. This would ensure that any changes made to the system would not compromise its overall security. Automating this process would also save time and reduce the risk of human error according to the participant.
 A third participant \changed{[C]} suggested the need for automating traceability among the different development artifacts, e.g., requirements and tests by automatically creating traceability links in a traceability model. This would allow developers to quickly identify which items need to be tested when new features or changes are introduced. It would also enable developers to track changes and updates more easily.
Another participant \changed{[E]} agreed that the link between threats and tests is crucial and suggested that it would be interesting to automate the process of indicating changes that need to be made by the developer in response to certain threats. This would help to ensure that changes are made quickly and accurately.

\emph{Threat modeling} is a critical aspect of any security strategy, but the process can be challenging, particularly when it comes to keeping threat libraries up to date: \interviewquote{we do not have a tool for example for the threat libraries.}{Case A} Automation can be used to make the threat modeling process more efficient and effective. \interviewquote{there could be ways to automate the threat models as well to catch the stuff from the code for example and pour it into the threat model.}{Case E}

Finally, one participant \changed{[F]} emphasized the need for a dashboard that provides a holistic overview of the whole system. Such a dashboard would enable monitoring the systems as a whole, as well as in a modular fashion.

\section{Discussion}

In this section, we discuss our results in relation to our research questions. We also provide key insights for each of the questions. 
Additionally, we discuss concerns related to evidence management that appeared in the results of multiple questions in a cross-cutting manner.

\subsection{RQ1: What is the context in safety-critical organizations \changed{with respect to} managing security evidence?}

It is evident from our results that the companies involved in this study are all aware of the importance of security evidence management. We believe that this also applies to other companies in safety-critical domains, as the main drivers of evidence work come from external sources such as regulations and standards.
However, the results indicate other internal needs for security assurance exist, which are relevant to working with security evidence. This aligns with the findings of \cite{mohamad2020security}. Lastly, a forcing driver is coming from customers which require security evidence for the products and services they are provided.

Our results indicate that there is a gap between the maturity of security evidence management processes and the level of maturity needed to fulfill the growing security demands in safety-critical domains such as automotive and medical.
We also noticed that bigger organizations have different maturity levels across the sub-organizations depending on the security cultures of different teams. 
This can be seen as an extended problem of inter-organizational security, which has been studied and reported in the literature. Karlsson et al. \cite{interOrg} report those in their systematic literature review and point out that the maturity level of the field is still low as most studies are descriptive, philosophical, or theoretical. 
However, within-organizational security has not been sufficiently studied, especially in the context of security evidence.
Hence, there is a need to further investigate within-organization security, which is needed to be able to have a security evidence management system for entire organizations.

\begin{mdframed}
\textbf{Key insight:}\\
There is awareness of the importance of managing security evidence. However, there is a big gap between the current maturity levels and those needed to cope with the growing requirements.
\end{mdframed}

\subsection{RQ2: How is the management of security evidence embedded in an organization's development process?}

During the software development process, many types of security evidence are created through various activities. These security artifacts play a crucial role in ensuring the security of software applications and data.

\changed{Security evidence is created throughout the development lifecycle. In early stages, initial risk analysis produces first potential risks and vulnerabilities. These artifacts are refined iteratively and risk mitigations are added as the system matures. In general, many of the evidence types mentioned by our interviewees are created in the later stages of the development lifecycle. For instance, in} testing and verification, the team tests the software application for vulnerabilities and ensures that it meets the specified security requirements. The evidence generated during this activity helps the team to ensure that the application is secure and meets the required standards.

\changed{Apart from this, more evidence is generated by other types of activities that are not tied to any specific development project or timeline, but rather to an organizational level.} One such activity is training. Training sessions can be conducted to educate employees about the importance of security and how to mitigate potential risks. The artifacts created during these training sessions can be considered essential security evidence, as they reflect the organization's commitment to ensuring security.

Despite their importance, security artifacts are not always considered special artifacts or evidence. Hence, there are no special processes to manage them, and they are managed as any other development artifact. This can lead to problems as security artifacts require special attention to ensure that they are appropriately managed and preserved. Without special processes in place, security artifacts may not be appropriately identified, documented, and retained, which introduces a risk of them not being utilized efficiently. Interestingly, a different observation has been reported by \citet{usman_rw}, where the study's participants stated that security was managed differently than other compliance requirements, and was considered to be a challenge. 
\changed{Many evidence types mentioned in this study (which we do not consider providing a complete list) are common with what is reported in the literature for safety~\cite{nair2014extended}. For example, the product information evidence types in \cite{nair2014extended} (code, verification and validation results, safety analysis results, etc.) are all mentioned and identified in this study. A notable absence is traceability information. Although there is a lack of a taxonomy to classify artifacts that can be considered evidence, the similarities in the results to these reported by \citet{nair2014extended} indicate the possibility for a knowledge transfer from safety evidence taxonomies to security or to create a combined safety and security taxonomy.}

\begin{mdframed}
\textbf{Key insight:}\\
Many security artifacts are created throughout the development process. However, they are currently not considered evidence and are thus not covered by specific processes to manage them. Rather, they are managed as any other development artifact.
\end{mdframed}

\subsection{RQ3: What are the detailed procedures aimed at managing security evidence in safety-critical organizations?}

The results indicate different ways in which the companies manage security evidence. Moreover, the different sub-organizations and teams also manage evidence differently. 
When it comes to roles and responsibilities, there are organizations that rely on development teams to carry out the evidence management work. Other organizations have more security-specialized roles that support the development teams. According to our results, it is desirable to have specialized roles, especially for the activities of owning, collecting, and governing security evidence. 
Our results also show that there are processes in place for evidence management. However, there are no processes that cover the entire scope of evidence management.

When it comes to storing and accessing evidence, we saw different opinions. There are two main strategies, which are centralized storage and decentralized. There is also a hybrid approach combining both strategies. All of these have advantages and disadvantages as discussed in the results. 

Structuring the evidence is not done systematically according to our results. In most cases, it follows the project or product structure at the company. In other cases, it is driven by organizational structures or the used tools.


Quality is a major issue. There are no concrete processes to ensure that the quality of the produced evidence is sufficient in a given context. It relies heavily on the confidence of the people producing the evidence. This gives a view of the human aspect which contrasts previous literature. The literature review conducted by \citet{humanFactor} found that the human factor is considered in the literature to be the main threat to the security of an organization’s assets and the main cause of security breaches. Whereas our results indicate that humans are the main source of confidence in security evidence. 

To summarize, managing evidence is done differently in different organizations due to aspects such as the size of the organization, the complexity of the products, and the compliance requirements. 
However, decisions have to be made in all of these different aspects discussed here. These decisions need to be driven by the usage scenarios of the evidence and the needs of its potential users.

\begin{figure}
    \centering
    \includegraphics[width=\linewidth]{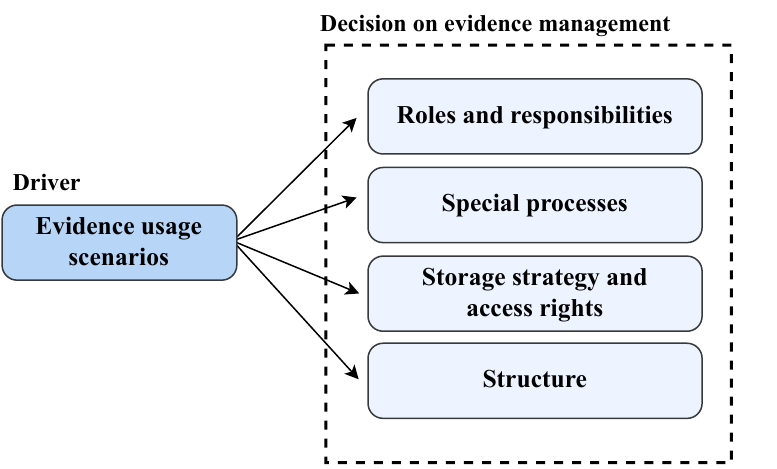}
    \caption{Security evidence management decisions and their drivers}
    \label{fig:management}
\end{figure}

Figure \ref{fig:management} illustrates the decisions that a company needs to make for managing security evidence and the main driver for these decisions. 

This can be a first step towards creating a framework for managing security evidence on an organizational level.

\begin{mdframed}
\textbf{Key insight:}\\
Companies carry out many activities to manage development security evidence on a team level, but there is a lack of an organizational-level framework to manage evidence.
\end{mdframed}

\subsection{RQ4: What are the challenges in managing security evidence for safety-critical organizations?}

After analyzing the results, it became clear that working with security evidence is a complex task that poses several challenges. Interestingly, most of these challenges are not technical but rather organizational in nature, while studies that tackle security compliance and evidence emphasize the technical challenges \cite{challenge_technical1,challenge_technical2}.

A significant challenge faced by the companies is scoping the work with security evidence. There is always a need to define an acceptable level of security to satisfy the requirements of customers and regulators. However, the regulations and standards do not set a clear scope for security evidence work and do not identify a specific security level to achieve.
\changed{This aligns with challenges found by \citet{ruiz2011challenges}, particularly the challenging task of determining the degree of compliance with standards for different domains and markets.}


However, the most significant challenge that companies face is finding competent personnel to work with security evidence and cybersecurity in general. This is because working with security evidence requires a specific skill set that is not easily found in the job market.
According to Furnel \cite{skills} the announcements for  cybersecurity roles do not capture the specific skills and qualifications needed for the role, but rather are very generic. We believe that this issue is emphasized for security evidence as the needed skills and qualifications are even more specific.

In addition to this, collaborating on a large scale between different teams to carry out security evidence work is challenging, particularly in the absence of processes. Addressing these organizational challenges is crucial for companies to ensure the security of their products and systems. However, it is not an easy task. In the long run, security will be more emphasized in education by universities which make security a part of the discipline of computer science \cite{training1,training_old}, as well as enterprises. This is due to the increasing awareness of the importance of security education \cite{training1,training2}, but companies have to act long before that. Hence, the way forward is to have more extensive and specialized security training for employees and try to establish a security culture.

\begin{mdframed}
\textbf{Key insight:}\\
The main challenges are organizational and related to structuring the work with security evidence and establishing the skill set rather than technical.
\end{mdframed}

\subsection{RQ5: To what extent can evidence management be supported through automation?}
There are multiple automated solutions applied in industries today that help work with security evidence, such as anomaly detection and various report generators. However, the needs are on a larger scale. There is a need to further automate activities to enable better management of security evidence, such as traceability, e.g., between different development artifacts and requirements, and also automating critical components such as threat modeling. 

One aspect that was mentioned by some participants is the potential usage of AI/machine learning to overcome issues related to the classification of artifacts and items. However, the participants did not suggest the use of AI to assist in tasks related to evidence management. For example, conversational AI systems such as ChatGPT \cite{chatGPT} and Github Copilot \cite{coPilot} have become very popular among practitioners in different fields as a tool for text generation and modification. In software engineering, there is a trend toward using AI in tasks related to code generation and documentation. Moreover, the AI can be used as a guide to the developers rather than a tool to generate content as discussed by \cite{melo2020}.
In security, in particular, a conversational AI can be used to, e.g., give a list of potential threats for a given system or platform. This can help practitioners to better control the quality of their implementation by considering items from the list. 
Most of the interviews in this study were conducted before ChatGPT was released. Other conversational AI systems were of course available, but these were not as popular as ChatGPT is today.
It is important to mention that using AI for software engineering tasks shall be taken cautiously. AI systems might produce inaccurate results. Moreover, they might create codes and texts that are under copyright. 


The use of machine learning for security-related tasks which are relevant to security evidence has been explored in the literature. For example, there are multiple studies to identify security requirements \cite{secReq1,secReq2,secReq3}, which was considered by our interviewees to be an important step for creating traceability links between the different security artifacts including the evidence. However, there is a need for further studies that expand this to other artifacts such as test cases and other types of evidence.

\begin{mdframed}
\textbf{Key insight:}\\
There are interesting ideas for automation, but practitioners do not yet understand the capabilities of AI and how it can help.
\end{mdframed}


\subsection{Human aspects}

\begin{figure}
    \centering
    \includegraphics[width=\linewidth]{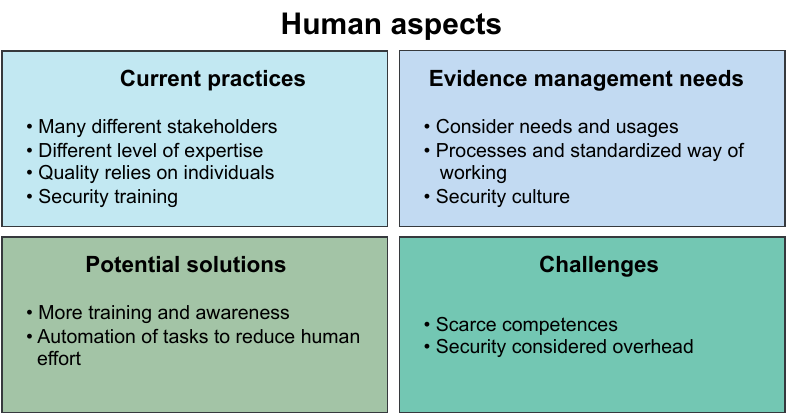}
    \caption{Human factors across multiple areas}
    \label{fig:desc_human}
\end{figure}

Upon analyzing our results, we found that the human aspect was a cross-cutting concern across all our research questions. Figure~\ref{fig:desc_human} illustrates the primary human-related points in four contexts: current practices, evidence management needs, challenges, and potential solutions.

In the current state of practice, security evidence has many stakeholders. These stakeholders have varying perspectives on how evidence should be managed driven by their needs and uses. In a certain company, the level of security expertise varies between sub-organizations and teams due to a lack of a consistent security culture throughout the company. The level of expertise typically correlates with the security awareness of the team's managers rather than the company's policy.

When it comes to measures for evidence quality assurance, there is a lack of processes and metrics in many cases. In such cases, the human aspect plays a crucial role. Evidence quality can only be assessed by considering the people involved in its creation. For instance, if experienced personnel create the evidence, it is considered high-quality evidence.
Furthermore, companies recognize the significance of security and provide regular security awareness drills as well as specific training to employees at different levels and scales. 
While the importance of security training is well understood, the aim should be to establish a security culture in organizations. A security culture reinforces the importance of security and helps employees integrate security practices into their daily work.
It also helps encourage employees to be proactive in identifying and reporting potential security risks. Moreover, a security culture empowers employees to take ownership of security within the company \cite{secCult}.

To manage security evidence effectively, it is essential to consider the different stakeholders, their distinct needs, and usage scenarios when it comes to evidence storage and structuring. For example, if many stakeholders require access to a specific type of evidence, it may be beneficial to store evidence of that type in a shared location that is easily accessible to all stakeholders. 
Additionally, collaboration among teams is necessary for evidence work. Therefore, it is vital to establish standardized working methods and processes that allow for communication and collaboration among teams.

However, challenges related to human aspects exist, including finding employees with security competencies, which is a significant hurdle as there is a far greater demand for security expertise than what the labor market can supply.
Another challenge related to human aspects is that the work related to security assurance, including evidence work, may be considered an overhead by some development team members, posing a risk that these tasks are left undone or conducted with low quality. 

To address these challenges, more security training and awareness must be conducted at companies. Such training should be employed to establish a security culture and enhance employees' competence regarding security \cite{humanFactor}. Additionally, to mitigate the risk of employees considering evidence work as an overhead, automation of tasks such as report generation can be applied.

\begin{mdframed}
\textbf{Key insights:}\\
Effective management of security evidence requires considering the human aspect, including stakeholders' needs and the human role in evidence quality assessment. 
\\
Establishing a security culture is crucial, but there are challenges such as finding employees with security competencies and addressing the perception that work with evidence is overhead.
\end{mdframed}

\subsection{Supplier-customer relation}

\begin{figure}
    \centering
    \vspace*{-2mm}
    \includegraphics[width=\linewidth]{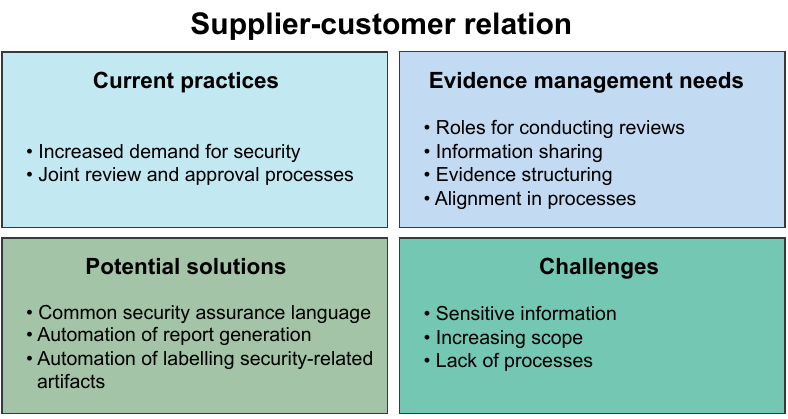}
    \caption{Supplier-customer relation across multiple areas}
    \label{fig:desc_supp}
\end{figure}

The supplier-customer relationship is another topic that was brought up in the results of multiple research questions.
Figure~\ref{fig:desc_supp} depicts the most important points related to supplier-customer relationships in four contexts, the current practices, the needs for evidence management, challenges, and potential solutions.

In the current state of practice, customers are requesting more security-related features and proof of security for their products. The regulations and standards that require security assurance and evidence in some industries, e.g., automotive, apply to bigger companies, e.g., OEMs. However, these companies rely on suppliers to provide different parts which are then integrated into the main product. Hence, the security requirements on the final product would be partially delegated to the suppliers as well.

To help achieve security assurance between suppliers and customers, it is common nowadays, according to our results, to conduct joint reviews and approval processes where the supplied artifacts are verified against the requirements. 

Having specialized roles for conducting these joint reviews is important. The reviewers have to possess enough knowledge about security and the current security trends to be able to assess the quality of the proposed security solutions. 
There is also a need to share a vast amount of information between the involved parties. This usually includes the customer and one supplier, but in special cases, it could also include multiple suppliers. 

In many cases, the evidence taken from a supplier would be integrated into a bigger artifact for the hosting product. Hence, it is important to \changed{ have a standardized set of terms, concepts, and frameworks to communicate and ensure a shared understanding of security assurance, i.e., a common language.}
It is also required to align the processes of the suppliers and customers to be able to fulfill evidence requirements. For example, a customer might need to ask for certain evidence in alignment with an internal process for security assessment. In that case, the customer has to ask for this evidence from the suppliers and they shall have the ability to provide it.

There are challenges when working between suppliers and customers. Firstly, there is an increased demand for security from customers that stretches out the scope of security at the supplier side both in their products and also their internal systems.
There is also a lack of processes that cover the relationship with suppliers in its entirety. This makes it hard to align processes between all involved parties. Moreover, evidence is generally considered sensitive data on many occasions, and hence sharing it with third parties becomes an issue. The problem multiplies when we have a triage relationship between a customer and multiple suppliers.

Mitigating these challenges can be supported by automating some tasks, e.g., the report generation of a security case, and automatically converting the assurance case into different formats.
Additionally, it would also be very beneficial to automate the labeling of security artifacts, e.g., requirements and test cases to make tasks such as joint reviews of security evidence easier and more manageable.

\begin{mdframed}
\textbf{Key insights:}\\
The supplier-customer relationship in the context of evidence management is increasingly complex. A common language for security assurance is key to coping with that, and automation can pave the way towards that.
\end{mdframed}
\section{Threats to validity}
\label{sec:threats}
In this section, we discuss the threats to the validity of this study based on the classification scheme provided by \citet{runeson12} and consider the threats to validity described by \citet{maxwell_validity}.

In terms of \textbf{Construct Validity}, we consider the risk of misunderstandings and misinterpretations by the participants of the \changed{interviews}, especially since we are referring to terms such as \emph{security evidence} which might not be clear to the participants that use different terminologies. Hence, we presented the purpose of the study and provided a brief context in the invitation emails sent to the participants. Moreover, we provided a definition and examples of the main concepts at the beginning of each \changed{interview}.

For \textbf{Internal Validity}, \changed{we consider the interviews themselves and the process of our thematic analysis.}

\changed{During the group interviews, we used moderation techniques to avoid the emergence of dominant voices by ensuring that all participants got their say. In the aftermath, we followed the advice by \citet{smithson2000using} to identify instances of dominant voices and normative discourses in the data, but did not find any.}

To mitigate the risk of descriptive validity threat, we recorded the \changed{interviews} to make sure that we documented the statements that the participants said in their original context rather than taking notes, which would risk being incomplete or misinterpreted by the reviewer. During the interviews, we tried to avoid reactivity by encouraging discussions among the participants whenever possible. 
After transcribing the sessions, we sent them back to the participants \changed{for member checking~\cite{candela2019exploring}} to further avoid any misunderstanding. 
When conducting the thematic coding, we mitigated the risk of subjective judgment by having two researchers code parts of the transcripts individually and then discussing all the instances where the codes differ to establish a baseline for the remainder of the coding process. For analyzing the codes and searching for patterns, all four researchers who authored this study collaborated in two workshops to avoid subjectivity and mitigate researcher bias.

In terms of \textbf{External Validity}, we are aware of the risk that the findings may not be applicable to all companies in safety-critical domains and markets due to the different regulations, standards, and best practices applied in the different domains. Additionally, the social aspect can differ between the companies in different parts of the world. To mitigate this, we conducted the study in six different companies which are active in three domains. Additionally, the companies are located in three different countries and have businesses worldwide.

We also considered the limited number of participants in our study. However, we aimed to have professionals with high expertise in safety-critical domains and are actively working with cybersecurity, which is a hard target. The majority of the participants have more than ten years of experience and three years of security experience. 

\changed{Our small sample size and the difficulty to find organisations with a sufficiently high maturity level also means that we cannot make claims about saturation. Since we consider this study exploratory in nature and our aim is to highlight the breadth of practices, we still believe that our results are useful and representative of the state of the practice.}

When it comes to \textbf{Reliability}, we consider the ability of other researchers to reproduce the study. For that, we make the interview guide and the codebook used in the thematic analysis available as supplemental material for future research.

\section{Conclusion}\label{sec:end}

The overall contribution of our work as laid out in the previous pages is an overview of the maturity w.r.t.\ managing security evidence in organizations that build safety-critical software systems. To achieve this contribution, we collected data from twelve security practitioners from six different case companies. We studied the state of practice of security evidence management, how the work of security evidence is embedded in an organization's development process, what detailed procedures exist to manage security evidence, what challenges exist in that context, and how automation can be applied to support practitioners in managing the evidence. 

Our results indicate that there is a lack of maturity in managing the growing requirements related to security evidence management. We also found that companies typically address the development of security evidence on a team level without an organizational-level framework. We found that the challenges associated with security evidence management are predominantly organizational rather than technical in nature. Moreover, we identified areas where automation could make evidence management more efficient.

Based on our findings, we plan to \changed{study and analyze existing tools and frameworks that can be used for security evidence management taking into consideration the needs and opportunities identified in this study.} \changed{Furthermore, we will use our findings to understand the gap between the state of practice and what is prescribed in current and upcoming security standards to identify the areas where research is still required.} In addition, we intend to explore how AI can help bridge the maturity gap and address the challenges associated with security evidence management.

\section*{Acknowledgement}
This work is partially supported by the CASUS research project funded
by VINNOVA, a Swedish funding agency. We sincerely thank the anonymous reviewers whose comments and suggestions helped improve and clarify this manuscript.

\bibliographystyle{model1-num-names}
\bibliography{references}

\appendix
\noindent
\section{Interview Questions}
\label{appx:a}

The interview questions along with their relation to the research questions can be found in Table~\ref{tab:interview-questions}.

\begin{table*}[h]
    \centering
    \caption{\changed{Interview questions and corresponding research questions}}
    \label{tab:interview-questions}
    \begin{tabular}{p{2.5cm}p{12cm}p{1.5cm}}
        \toprule
        & \textbf{\changed{Question}} & \textbf{\changed{RQ}} \\
        \midrule 
         \multirow{14}{*}{\changed{State of practice}}  & - \changed{Starting off with the company: what products/services do you provide and who are the main customers?} & \multirow{14}{1.5cm}{\changed{RQ1, RQ2, RQ4}} \\
         & - \changed{How important is the security aspect when it comes to the overall production / service providing?} & \\
         & - \changed{In high-level terms, how is security assurance addressed in the company?} & \\
         & - \changed{What is the main motivation for security work in the company? Is it an external driver (standards, regulations… etc) or are there other internal drivers?} & \\
         & - \changed{Do you create SAC for your products/systems?} & \\
         & - \changed{What is a security evidence for you? Follow up: to us, a security evidence is any artifact (test case, test report, document... etc) that contributes towards the overall confidence in the security of the system in question} &  \\
        
         & - \changed{What kinds of security evidence exist at the company?} & \\
         & - \changed{In particular, what evidence exist in each of the following categories: *Here we show the BSIMM framework*. Could you give some examples?} & \\
         \midrule
         \multirow{4}{*}{\changed{Responsibility}}  & - \changed{Who is responsible for producing evidence?} & \multirow{4}{*}{\changed{RQ3}} \\
         & - \changed{Who is responsible for maintaining evidence?} & \\
         & - \changed{Are there specific roles for managing / working with security evidence?} & \\
         & - \changed{Who are the consumers/users of this evidence? (engineers, SAC creators... etc)?} & \\
         \midrule
         \multirow{4}{*}{\changed{Logistics}}  & - \changed{Where is the evidence usually stored?} & \multirow{4}{*}{\changed{RQ3}} \\
         & - \changed{How is the access to this evidence given?} & \\
         & - \changed{How is the evidence structured? Is the evidence associated with features or products?} & \\
         & - \changed{What processes exists to manage evidence? What should be there?} & \\
         \midrule
         \multirow{8}{*}{\changed{Technical Aspects}}  & - \changed{Why do you collect these evidence? Can you give some examples of use-cases?} & \multirow{8}{*}{\changed{RQ3, RQ4}} \\
         & - \changed{Can you think of types of evidence (that do not exist today) that would be necessary for current or future use-cases?} & \\
         & - \changed{When is the evidence created? (In which activity/processes step)} & \\
         & - \changed{How is the evidence traced to security issues / claims?} & \\
         & - \changed{How is the evidence maintained?} & \\
         & - \changed{What kind of properties does the evidence have? E.g., confidence, sensitivity... etc. And how do these properties affect / should affect the management of the evidence?} & \\
         \midrule
         \multirow{2}{*}{\changed{Automation}}  & - \changed{Are there any parts of the evidence management that is automated? In that case which?}  & \multirow{2}{*}{\changed{RQ 5}} \\
         & - \changed{What do you think should be automated? Please start with the most important ones.} & \\
         \bottomrule
    \end{tabular}
\end{table*}

\end{document}